\documentclass[12pt,a4paper,twoside]{article}
\usepackage{amsmath,amssymb,amsfonts,amscd}
\usepackage{cancel}
\usepackage{mathtools}
\usepackage{theorem}
\usepackage{color}
\usepackage{graphicx}
\usepackage{mdframed}
\usepackage{epsfig}
\usepackage{helvet}
\usepackage{verbatim}
\usepackage{pict2e}

\newcommand{\R}{{\mathord{\mathbb R}}}
\newcommand{\Z}{{\mathord{\mathbb Z}}}
\newcommand{\N}{{\mathord{\mathbb N}}}
\newcommand{\C}{{\mathord{\mathbb C}}}

\newcommand{\mL}{{\mathcal L}}

\newcommand{\mO}{{\mathcal O}}

\renewcommand{\Re}{{\rm Re}}
\renewcommand{\Im}{{\rm Im}}
\newcommand{\e}{{\rm e}}

\renewcommand{\i}{{\rm i}}

\newcommand{\ran}{{\rm ran\,}}

\newcommand{\tr}{{\rm tr}}

\def\slim{\mathop{\rm s-lim}}

\newcommand{\Aut}{{\rm Aut}}

\newcommand{\rd}{{\rm d}}

\newcommand{\Mat}{{\rm Mat}}

\newcommand{\fh}{{{\mathfrak h}}}

\newcommand{\ff}{{{\mathfrak f}}}

\newcommand{\fA}{{\mathfrak A}}

\newcommand{\eps}{{\epsilon}}

\newcommand{\spec}{{\rm spec}}

\newcommand{\vi}{\varphi}

\newcommand{\gam}{\gamma}

\newtheorem{thm}{Theorem}
\newtheorem{proposition}[thm]{Proposition}
\newtheorem{lemma}[thm]{Lemma}
\newtheorem{definition}[thm]{Definition}
{\theorembodyfont{\upshape} \newtheorem{remark}[thm]{\it Remark}}
\newtheorem{example}[thm]{Example}
\newtheorem{assumption}[thm]{Assumption}
\newtheorem{setting}[thm]{Setting}
\newtheorem{corollary}[thm]{Corollary}
\newcommand{\bd}{\begin{definition}}
\newcommand{\ed}{\end{definition}\vspace{1mm}}
\newcommand{\bt}{\begin{thm}}\vspace{1mm}
\newcommand{\et}{\end{thm}}
\newcommand{\bc}{\begin{corollary}}
\newcommand{\ec}{\end{corollary}\vspace{1mm}}
\newcommand{\bl}{\begin{lemma}}
\newcommand{\el}{\end{lemma}\vspace{1mm}}
\newcommand{\bp}{\begin{proposition}}
\newcommand{\ep}{\end{proposition}\vspace{1mm}}
\newcommand{\bx}{\begin{example}}
\newcommand{\ex}{\end{example}\vspace{1mm}}
\newcommand{\br}{\begin{remark}}
\newcommand{\er}{\end{remark}}
\newcommand{\bass}{\begin{assumption}}
\newcommand{\eass}{\end{assumption}\vspace{1mm}}
\newcommand{\bset}{\begin{setting}}
\newcommand{\eset}{\end{setting}\vspace{1mm}}
\newcommand{\bprf}{\noindent{\it Proof.}\, }
\newcommand{\eprf}{\hfill $\Box$\vspace{5mm}}

\newcommand{\ie}{\noindent{i.e.}}

\def\bas#1\eas{\begin{align*}#1\end{align*}}
\def\ba#1\ea{\begin{align}#1\end{align}}

\newcommand{\bn}{\begin{enumerate}}
\newcommand{\en}{\end{enumerate}}
\newcommand{\ita}{\item[\it (a)]}
\newcommand{\itb}{\item[\it (b)]}

\newcommand{\lm}{\lambda}
\renewcommand{\d}{{\rm d}}

\begin{document}
\pagestyle{myheadings}
\markboth{Walter H. Aschbacher}{On a quantum phase transition in a steady state out of equilibrium}

\title{On a quantum phase transition in a steady state out of equilibrium}

\author{Walter H. Aschbacher\footnote{http://aschbacher.univ-tln.fr}
\\ \\
Aix Marseille Universit\'e, CNRS, CPT, UMR 7332, 13288 Marseille, France\\
Universit\'e de Toulon, CNRS, CPT, UMR 7332, 83957 La Garde, France 
}

\date{}
\maketitle
\begin{abstract}
Within the rigorous axiomatic framework for the description of quantum
mechanical systems with a large number of degrees of freedom, we show that
the nonequilibrium steady state, constructed in the quasifree fermionic
system corresponding to the isotropic XY chain in which a finite sample,
coupled to two thermal reservoirs at different temperatures, is exposed to 
a local external magnetic field, is breaking translation invariance and
exhibits a strictly positive entropy production rate. Moreover, we prove 
that there exists a second-order nonequilibrium quantum phase transition 
with respect to the strength of the magnetic field as soon as the system
is truly out of equilibrium.
\end{abstract}


\section{Introduction}

{\it "As useful as the characterization of equilibrium states by 
thermostatic theory has proven to be, it must be conceded that our primary
interest is frequently in processes rather than in states. In biology,
particularly, it is the life process that captures our imagination
rather than the eventual equiilbrium} [sic] {\it state to which each 
organism inevitably proceeds."}\\
\hspace*{\fill}H. B. Callen \cite[p.\hspace{0.4mm}283]{HBC}

\vspace{5mm}

A precise analysis of quantum mechanical systems having a large, \ie, often, 
in physically idealized terms, an infinite number of degrees of freedom is
most effectively carried out within the mathematically rigorous framework
of operator algebras. As a matter of fact, having been heavily used in
the 1960s for the description of quantum mechanical systems in thermal
equilibrium, the benefits of this framework have again started to unfold 
more recently in the physically much more general situation of open quantum
systems out of equilibrium. In the latter field, vast by its very nature,
most of the mathematically rigorous results have been obtained for a
particular family of states out of equilibrium, namely for the so-called 
nonequilibrium steady states (NESS) introduced in 
\cite[p.\hspace{0.4mm}6]{R} as the large time limit of the averaged 
trajectory of some initial state along the full time evolution. Beyond the
challenging construction of this central object of interest in physically
interesting situations, the derivation, from first principles, of 
fundamental transport and quantum phase transition properties for
thermodynamically nontrivial systems out of equilibrium has also come
within reach. 

In both quantum statistical mechanics in and out of equilibrium, an 
important role is played by the so-called quasifree fermionic systems, 
and this is true not only with respect to the mathematical accessibility
but also when it comes to real physical applications. Namely, from a 
mathematical point of view, such systems allow for a simple and powerful
description by means of scattering theory restricted to the underlying 
one-particle Hilbert space over which the fermionic algebra of observables
is constructed. This restriction of the dynamics to the one-particle 
Hilbert space opens the way for a rigorous mathematical 
analysis of many properties which are of fundamental physical interest.
On the other hand, they also constitute a class of systems which are 
indeed realized in nature. One of the most prominent representatives of 
this class is the so-called XY spin chain introduced in 1961 in 
\cite[p.\hspace{0.4mm}409]{LSM} where a specific equivalence of this spin model with a quasifree fermionic system has been established. Already in
the 1960s, the first real physical candidate has been identified in 
\cite[p.\hspace{0.4mm}459]{CSP} (see, for example, \cite{MK} for a survey
of the rich interplay between the experimental and theoretical research 
activity in the very dynamic field of low-dimensional magnetic systems).

In the present paper, we analyze transport and quantum phase transition
properties for the quasifree fermionic system over the two-sided discrete
line corresponding, in the spin picture, to the special case of the 
so-called isotropic XY spin chain. In order to model the desired
nonequilibrium situation as in \cite[p.\hspace{0.4mm}3431]{A}, we first 
fix 
\ba
\nu\in\N_0,
\ea
and cut the finite piece of length $2\nu+1$, 
\ba
\label{ZS}
\Z_S
:=\{x\in\Z\,|\, |x|\le \nu\},
\ea
out of the two-sided discrete line. This piece plays the role of the 
configuration space of the confined sample, whereas the remaining parts,
\ba
\label{ZL}
\Z_L
&:=\{x\in\Z\,|\, x\le -(\nu+1)\},\\
\label{ZR}
\Z_R
&:=\{x\in\Z\,|\, x\ge \nu+1\},
\ea
act as the configuration spaces of the infinitely extended thermal
reservoirs. Furthermore, over these configuration spaces, an initial decoupled state is
prepared as the product of three thermal equilibrium states carrying the
corresponding inverse temperatures
\ba
\label{temp}
0
=\beta_S
<\beta_L
<\beta_R
<\infty.
\ea
Finally, the NESS is constructed with respect to the full time evolution
which not only recouples the sample and the reservoirs but also introduces a
one-site magnetic field of strength 
\ba
\label{mag}
\lm\in\R
\ea
at the origin.

As a first result, we show that there exists a unique quasifree NESS,
which we call the magnetic NESS, whose translation invariance is broken 
as soon as the local magnetic field strength $\lm$ is switched on. This
property contrasts with the translation invariance of the NESS constructed
in \cite[p.\hspace{0.4mm}1158]{AP} for the anisotropic XY model with 
spatially constant magnetic field (see also Remark \ref{rem:xy} below).
Moreover, we explicitly determine the NESS expectation value of the 
extensive energy current observable describing the energy flow through 
the sample from the left to the right reservoir. The form of this current
immediately yields that the entropy production rate, the first fundamental
physical quantity for systems out of equilibrium, is strictly positive. 
Finally, we show that the system also exhibits what we call a second-order
nonequilibrium quantum phase transition, \ie, in the case at hand, a
logarithmic divergence of the second derivative of the entropy production 
rate with respect to the magnetic field strength at the origin.

The present paper is organized as follows.\\
{\it Section \ref{sec:neqs}}\, specifies the nonequilibrium setting, \ie,
it introduces the observables, the quasifree dynamics, and the quasifree
initial state.\\
{\it Section \ref{sec:ness}}\, contains the construction of the magnetic 
NESS and the proof of the breaking of translation invariance for 
nonvanishing magnetic field strength.\\
{\it Section \ref{sec:ep}}\, introduces the notions of energy current and
entropy production rate. It also contains the determination of the heat
flow and the strict positivity of the entropy production rate.\\
{\it Section \ref{sec:nqpt}}\, displays the last result, \ie, the proof of 
the existence of a second-order nonequilibrium quantum phase transition.\\
Finally, the {\it Appendix}\, collects some ingredients used in the 
foregoing sections pertaining to the spectral theory of the magnetic
Hamiltonian, the structure of the XY NESS, and the action of the magnetic 
wave operators on completely localized wave functions.

\section{Nonequilibrium setting}
\label{sec:neqs}

Remember that, in the operator algebraic approach to quantum
statistical mechanics, a physical system is specified by an algebra of
observables, a group of time evolution automorphisms, and a normalized
positive linear state functional on this observable algebra (see, for
example, the standard references \cite{BR} for more details). 

In the Definition \ref{def:obs}, \ref{def:qfdyn}, and \ref{def:qfs} below,
part {\it (a)} recalls the general formulation for quasifree fermionic
systems and part {\it (b)} specializes to the concrete situation from 
\cite[p.\hspace{0.4mm}3431]{A} which we are interested in in the present
paper.

In the following, the commutator and the anticommutator of two elements
$a$ and $b$ in the corresponding sets in question read as usual as
 $[a,b]:=ab-ba$ and $\{a,b\}:=ab+ba$, respectively.

\bd[Observables]
\label{def:obs}
\hspace{0mm}
\vspace{-3mm}
\bn
\ita
Let $\fh$ be the one-particle Hilbert space of the physical system. 
The observables are described by the elements of the unital canonical anticommutation algebra $\fA$ over $\fh$ 
generated by the identity $1$ and elements $a(f)$ for all $f\in\fh$, 
where $a(f)$ is antilinear in $f$ and, for all $f,g\in\fh$, we have the
canonical anticommutation relations
\ba
\label{car1}
\{a(f), a(g)\}
&=0,\\
\label{car2}
\{a(f), a^\ast(g)\}
&=(f,g)\, 1,
\ea
where 
$(\hspace{0.3mm}\cdot\hspace{0.5mm},\hspace{0.2mm}\cdot\hspace{0.3mm})$
denotes the scalar product of $\fh$.

\itb
Let the configuration space of the physical system be given by the 
two-sided infinite discrete line $\Z$ and let 
\ba
\fh
:=\ell^2(\Z)
\ea
be the one-particle Hilbert space over this configuration space. 
\en
\ed

The second ingredient is the time evolution. In order to define a 
quasifree dynamics, it is sufficient, by definition, to specify its 
one-particle Hamiltonian, i.e., the generator of the time evolution on the
one-particle Hilbert space $\fh$. In general, a one-particle Hamiltonian 
is an unbounded selfadjoint operator on $\fh$ (which is typically bounded
from below though). Such an action on the one-particle Hilbert space can 
be naturally lifted to the observable algebra $\fA$, becoming a group of
automorphisms of $\fA$. 

In the following, the set of ($\ast$-)automorphisms of $\fA$ will be
written as $\Aut(\fA)$ and we denote by $\mL(\fh)$ and $\mL^0(\fh)$
the bounded operators and the finite rank operators on $\fh$, 
respectively. Moreover, for all $r,q\in\R$, the Kronecker symbol is 
defined as usual by $\delta_{rq}:=1$ if $r=q$ and $\delta_{rq}:=0$ if
$r\neq q$. Furthermore, for all $a\in\mL(\fh)$, we define the real part 
and the imaginary part of $a$ by $\Re(a):=(a+a^\ast)/2$ and 
$\Im(a):=(a-a^\ast)/(2\i)$, respectively. Finally, we set $a^0:=1\in\mL(\fh)$, and,
if $a$ is invertible, we use the notation $a^{-n}:={(a^{-1})}^n$ for all 
$n\in\N$.

\bd[Quasifree dynamics]
\label{def:qfdyn}
\hspace{0mm}
\vspace{-3mm}
\bn
\ita
Let $h$ be a Hamiltonian on $\fh$. The quasifree dynamics generated by
$h$ is the automorphism group defined, for all $t\in\R$ and all $f\in\fh$,
by 
\ba
\label{auto}
\tau^t(a(f))
&:=a(\e^{\i t h}f),
\ea
and suitably extended to the whole of $\fA$. 

\itb
The right translation $u\in\mL(\fh)$ and the localization projection 
$p_0\in\mL^0(\fh)$ are defined, for all $f\in\fh$ and all $x\in\Z$, by 
\ba
\label{u}
(uf)(x)
&:=f(x-1),\\
p_0f
&:= f(0)\hspace{0.5mm}\delta_0,
\ea
where the elements of the usual completely localized orthonormal Kronecker
basis $\{\delta_y\}_{y\in\Z}$ of $\fh$ are given by 
$\delta_y(x):=\delta_{xy}$ for all $x\in\Z$. Moreover, let $\lm\in\R$ 
denote the magnetic field strength and define the 
one-particle Hamiltonians $h,h_{\rm d}, h_\lm\in\mL(\fh)$ by
\ba
\label{h}
h
&:=\Re(u),\\
\label{hd}
h_{\rm d}
&:=h-(v_L+v_R),\\
\label{hlm}
h_\lm
&:=h+\lm v,
\ea
where the local external magnetic field $v\in \mL^0(\fh)$ and the 
decoupling operators  $v_L,v_R\in \mL^0(\fh)$ read
\ba
v
&:=p_0,\\
v_L
&:=\Re(u^{-\nu}p_0u^{\nu+1}),\\
v_R
&:=\Re(u^{\nu+1} p_0u^{-\nu}).
\ea
The operators $h$, $h_{\rm d}$, and $h_\lm$ will be called the XY 
Hamiltonian, the decoupled Hamiltonian, and the magnetic Hamiltonian,
respectively. The corresponding time evolution automorphisms of $\fA$ are
given by \eqref{auto} and will be denoted by $\tau^t$, $\tau^t_\d$, and 
$\tau_\lm^t$ for all $t\in\R$.
\en
\ed

\br
\label{rem:xy}
As mentioned in the Introduction, the model specified by Definition
\ref{def:obs} and \ref{def:qfdyn} has its origin in the XY spin chain 
whose formal Hamiltonian reads
\ba
\label{H-XY}
H
=-\frac{1}{4}\sum_{x\in\Z}\left\{(1+\gamma)\,\sigma_1^{(x)}
\sigma_1^{(x+1)}+(1-\gamma)\,\sigma_2^{(x)}\sigma_2^{(x+1)}
+2\mu\hspace{0.3mm}\sigma_3^{(x)}\right\},
\ea
where $\gamma\in(-1,1)$ represents the anisotropy, $\mu\in\R$ the 
spatially homogeneous magnetic field, and 
$\sigma_1, \sigma_2, \sigma_3$ are the usual Pauli matrices. Indeed, 
using the so-called Araki-Jordan-Wigner transformation introduced in
\cite[p.\hspace{0.4mm}279]{Araki84} (see also Remark 
\ref{rem:araki}), the Hamiltonian from \eqref{h} 
corresponds to the case of the so-called isotropic XY chain without 
magnetic field, \ie, to the case with $\gamma=\mu=0$ in \eqref{H-XY}.
In order to treat the anisotropic case $\gam\neq 0$, one uses the 
so-called selfdual quasifree setting developed in 
\cite[p.\hspace{0.4mm}386]{Araki71}. In this most natural framework, one
works in the doubled one-particle Hilbert space $\fh\oplus\fh$ and the
generator of the truly anisotropic XY dynamics has nontrivial off-diagonal
blocks on $\fh\oplus\fh$ (which vanish for $\gam=0$). In many respects, 
the truly anisotropic XY model is substantially more complicated than the 
isotropic one (this is true, a fortiori, if the magnetic field $\mu$ is 
switched on whose contribution to the generator acts diagonally on 
$\fh\oplus\fh$ though). 
\er

\br
For all $\alpha\in\{L,S,R\}$, let us define the Hilbert space
\ba
\fh_\alpha
:=\ell^2(\Z_\alpha)
\ea
over the sample and reservoir configuration spaces $\Z_\alpha$ defined in
\eqref{ZS}, \eqref{ZL}, and \eqref{ZR}, respectively. Moreover, for all 
$\alpha\in\{L,S,R\}$, we define the map $i_\alpha^{}:\fh_\alpha\to\fh$, 
for all $f\in\fh_\alpha$ and all $x\in\Z$, by
\ba
(i_\alpha^{} f)(x)
:=\begin{cases}
f(x), & x\in\Z_\alpha,\\
0, & x\in\Z\setminus\Z_\alpha,
\end{cases}
\ea
and we note that its adjoint $i_\alpha^\ast:\fh\to\fh_\alpha$ is the 
operator acting by restriction to $\fh_\alpha$. With the help of these
natural injections, we define the unitary operator 
$w:\fh\to\fh_L\oplus\fh_S\oplus\fh_R$, for all $f\in\fh$, by
\ba
\label{vi}
w f
:=i_L^\ast f\oplus i_S^\ast f\oplus i_R^\ast f,
\ea
and we observe that its inverse $w^{-1}:\fh_L\oplus\fh_S\oplus\fh_R\to\fh$ 
is given by $w^{-1}(f_L\oplus f_S\oplus f_R)=i_Lf_L+i_Sf_S+i_Rf_R$ for all
$f_L\oplus f_S\oplus f_R\in\fh_L\oplus\fh_S\oplus\fh_R$. Then, using
\eqref{vi}, we can write
\ba
\label{wvL}
w v_L w^{-1}
&=\frac12\begin{bmatrix}
0 & (i_S^\ast\delta_{-\nu},\cdot\,)\hspace{0.3mm}i_L^\ast\delta_{-(\nu+1)} & 0\\
(i_L^\ast\delta_{-(\nu+1)},\cdot\,)\hspace{0.3mm}i_S^\ast\delta_{-\nu} & 0 & 0\\
0 & 0 & 0
\end{bmatrix},\\
\label{wvR}
w v_R w^{-1}
&=\frac12\begin{bmatrix}
0 & 0 & 0\\
0 & 0 & (i_R^\ast\delta_{\nu+1},\cdot\,)\hspace{0.3mm}i_S^\ast\delta_\nu\\
0 & (i_S^\ast\delta_\nu,\cdot\,)\hspace{0.3mm}i_R^\ast\delta_{\nu+1} & 0
\end{bmatrix},
\ea
where we used the same notation for the scalar products on $\fh_L$, 
$\fh_S$, and $\fh_R$ and for the one on $\fh$. Computing $w hw^{-1}$ and
using \eqref{wvL} and  \eqref{wvR}, we find that the decoupling operators 
$v_L$ and $v_R$ indeed decouple the sample from the reservoirs in the 
sense that
\ba
\label{whd}
w h_{ \rm d}w^{-1}
=h_L\oplus h_S\oplus h_R,
\ea
where, for all $\alpha\in\{L,S,R\}$, the Hamiltonians 
$h_\alpha\in\mL(\fh_\alpha)$ are defined by
\ba
h_\alpha
:=i_\alpha^\ast h i_\alpha^{}.
\ea
\er

\vspace{3mm}

The last ingredient are the states, \ie, the normalized positive linear
functionals on the observable algebra $\fA$. As discussed in the
Introduction, we are interested in quasifree states, \ie, in states whose
many-point correlation functions factorize in determinantal form.

In the following, $E_\fA$ stands for the set of states on $\fA$.
Moreover, for all $n\in\N$, the $n\times n$ matrix with entries $a_{ij}\in\C$ 
for all $i,j\in\{1,\ldots,n\}$ is denoted by $[a_{ij}]_{i,j=1}^n$.

\newpage

\bd[Quasifree states]
\label{def:qfs}
\hspace{0mm}
\vspace{-3mm}
\bn
\ita
Let $s\in\mL(\fh)$ be an operator satisfying $0\le s\le 1$.
A state $\omega_s\in E_\fA$ is called a quasifree state induced by $s$
if, for all $n,m\in\N$, all $f_1,\ldots, f_n\in\fh$, and all 
$g_1,\ldots, g_m\in\fh$, we have
\ba
\label{qf}
\omega_s(a^\ast(f_n)\ldots a^\ast(f_1)a(g_1)\ldots a(g_m))
=\delta_{nm} \det([(g_i,s f_j)]_{i,j=1}^n).
\ea
The operator $s$ is called the two-point operator of the quasifree
state $\omega_s$.

\itb
The decoupled initial state is defined to be the quasifree state 
$\omega_\d\in E_\fA$ induced by the two-point operator $s_\d\in\mL(\fh)$ 
of the form
\ba
\label{sd}
s_\d
:=\rho(\beta_L i_L h_Li_L^\ast+\beta_R i_Rh_Ri_R^\ast),
\ea
where, for all $r\in\R$, the Planck density function 
$\rho_r:\R\to\R$ is defined, for all $e\in\R$, by
\ba
\label{rhor}
\rho_r(e)
:=\frac{1}{1+\e^{r e}},
\ea
and, in \eqref{sd}, we used the simplified notation $\rho:=\rho_1$.
\en
\ed

\br
\label{rem:sd}
In the selfdual quasifree setting mentioned in Remark \ref{rem:xy}, the 
quasifreeness of a state is expressed by means of a pfaffian of the
two-point correlation matrix with respect to the generators of the selfdual canonical anticommutation algebra (see, for example, \cite[p.\hspace{0.4mm}3447]{A}).
\er

\br
Using the unitarity of $w\in\mL(\fh,\fh_L\oplus\fh_S\oplus\fh_R)$ from 
\eqref{vi}, the unitary invariance of the spectrum, the uniqueness of
the continuous functional calculus, and 
\ba
w\hspace{0.4mm} i_L h_Li_L^\ast w^{-1}
&= h_L\oplus 0\oplus 0,\\
w\hspace{0.4mm}i_R h_Ri_R^\ast w^{-1}
&=0\oplus 0\oplus h_R,
\ea
the two-point operator of the decoupled quasifree initial state can be written as
\ba
\label{wsd}
w s_\d w^{-1}
&=w\rho(\beta_L i_L h_Li_L^\ast+\beta_R i_Rh_Ri_R^\ast)w^{-1}\nonumber\\
&=\rho(w(\beta_L i_L h_Li_L^\ast+\beta_R i_Rh_Ri_R^\ast)w^{-1})\nonumber\\
&=\rho(\beta_L h_L\oplus 0\oplus \beta_R h_R)\nonumber\\
&=\rho_{\beta_L}(h_L)\oplus \tfrac121\oplus \rho_{\beta_R}(h_R),
\ea
\ie, the two-point operator $s_\rd$ does not couple the sample and the
reservoir subsystems as required by our nonequilibrium setting.
\er

One of the main motivations for introducing the magnetic Hamiltonian 
\eqref{hlm} in \cite[p.\hspace{0.4mm}3432]{A} was to study the effect of 
the breaking of translation invariance. For quasifree states as given in
Definition \ref{def:qfs} {\it (a)}, this property can be defined as follows.

\bd[Translation invariance]
\label{def:ti}
A quasifree state $\omega_s\in E_\fA$ induced by the two-point operator
$s\in\mL(\fh)$ is called translation invariant if 
\ba
[s,u]
=0, 
\ea
where $u\in\mL(\fh)$ is the right translation from \eqref{u}.
\ed

\br
By using the translation automorphism $\tau_u\in\Aut(\fA)$ defined by
$\tau_u(a(f)):=a(uf)$ for all $f\in\fh$ and suitably extended to the
whole of $\fA$, we note that a quasifree state $\omega_s\in E_\fA$ is translation invariant if and only if $\omega_s\circ\tau_u=\omega_s$.
\er

\section{Nonequilibrium steady state}
\label{sec:ness}

The definition of a nonequilibrium steady state (NESS) below stems
from \cite[p.\hspace{0.4mm}6]{R}. We immediately specialize it to the 
case at hand.

In the following, if nothing else is explicitly stated, we will always use 
the assumptions $0=\beta_S<\beta_L<\beta_R<\infty$ and $\lm\in\R$ from
\eqref{temp} and \eqref{mag}. Moreover, we will also use the notation
\ba
\label{delta}
\delta
&:=\frac{\beta_R-\beta_L}{2},\\
\label{beta}
\beta
&:=\frac{\beta_R+\beta_L}{2}.
\ea

\bd[Magnetic NESS]
\label{def:ness}
The state $\omega_\lm\in E_\fA$, defined, for all $A\in\fA$, by
\ba
\label{mness}
\omega_\lm(A):=
\lim_{t\to\infty}\hspace{1mm}\omega_\d(\tau_\lm^t(A)),
\ea
is called the magnetic NESS associated with the decoupled quasifree 
initial state $\omega_\d\in E_\fA$ and the magnetic dynamics 
$\tau_\lm^t\in\Aut(\fA)$ for all $t\in\R$.
\ed

\br
The more general definition in \cite[p.\hspace{0.4mm}6]{R} defines a NESS
as a weak-$\ast$ limit point for $T\to\infty$ of 
\ba
\frac1T\int_0^T\rd t\,\, \omega_\d\circ\tau_\lm^t.
\ea
This averaging procedure implies, in general, the existence of a limit 
point. In particular, in the quasifree setting, it allows to treat a 
nonvanishing contribution to the point spectrum of the one-particle 
Hamiltonian generating the full time evolution. In the present case, 
since, due to Proposition \ref{prop:spctr} {\it (c)} from Appendix 
\ref{app:hlm}, the Hamiltonian $h_\lm$ has a single eigenvalue, 
\eqref{mness} is sufficient to extract a limit from the point spectrum 
subspace (see \eqref{contr-pp} below).
\er

In \cite[p.\hspace{0.4mm}3436]{A}, we found that the magnetic NESS is a 
quasifree state induced by a two-point operator $s_\lm\in\mL(\fh)$ whose 
form could be determined with the help of scattering theory on the 
one-particle Hilbert space $\fh$. In particular, we made use of the wave 
operator $w(h_\d,h_\lm)\in\mL(\fh)$ defined by
\ba
\label{wave}
w(h_\d,h_\lm)
:=\slim_{t\to\infty}\e^{-\i t h_\d} \e^{\i t h_\lm} 1_{\rm ac}(h_\lm),
\ea
where $1_{\rm ac}(h_\lm)\in\mL(\fh)$ denotes the spectral projection onto
the absolutely continuous subspace of the magnetic Hamiltonian $h_\lm$.
Moreover, we will also need the spectral projection 
$1_{\rm pp}(h_\lm)\in\mL(\fh)$ onto the pure point subspace of $h_\lm$. 
All the spectral properties of the magnetic Hamiltonian $h_\lm$ which we 
will use in the following are summarized in Appendix \ref{app:hlm}.

\bt[Magnetic two-point operator]
\label{thm:ness-2}
The magnetic NESS $\omega_\lm$ is the quasifree state induced by the 
two-point operator $s_\lm\in\mL(\fh)$ given by
\ba
\label{slm}
s_\lm
=w^\ast(h_\d,h_\lm) s_\d w(h_\d,h_\lm)
+1_{\rm pp}(h_\lm) s_\d 1_{\rm pp}(h_\lm).
\ea
Moreover, if $\lambda\neq 0$, the pure point subspace of $h_\lm\in\mL(\fh)$ is one-dimensional.
\et

\br
\label{rem:xyness}
For all $\gamma\in (-1,1)$ and all $\mu\in\R$ in \eqref{H-XY}, the 
so-called XY NESS has been constructed in 
\cite[p.\hspace{0.4mm}1170]{AP} using time dependent scattering theory 
(see Theorem \ref{thm:xyness} of Appendix \ref{app:xyness} for the XY 
NESS with  $\gamma=\mu=0$). The two-point operator of the XY NESS for 
$\gamma=\mu=0$ coincides with the two-point operator of the magnetic NESS 
for $\lm=0$.
\er

\br
\label{rem:araki}
For the special case $\gamma=\mu=0$, the XY NESS has also been found in 
\cite{AH} with the help of asymptotic approximation methods. The 
construction of the XY NESS in \cite{AH} and \cite{AP} is carried out 
within the mathematically rigorous axiomatic framework of operator 
algebras for the description of quantum mechanical systems having an 
infinite number of degrees of freedom. Due to the two-sidedness of the 
present nonequilibrium setting, the passage from the spin algebra to the 
canonical anticommutation algebra relies on the so-called 
Araki-Jordan-Wigner transformation introduced in 
\cite[p.\hspace{0.4mm}279]{Araki84}. In contrast to the case of the usual 
Jordan-Wigner transformation for finite or infinite one-sided systems, the 
direct correspondence between these two algebras breaks down in the 
thermodynamic limit of an infinite chain which extends in both directions. 
\er

\br
\label{rem:approaches}
In \cite[p.\hspace{0.4mm}1158]{AP}, we observed that the XY NESS can be
written, formally, as an equilibrium state specified by the inverse
temperature $\beta$ and an effective Hamiltonian which differs from the
original XY Hamiltonian by conserved long-range multi-body charges (in 
\cite[p.\hspace{0.4mm}914]{MO}, it has been proved though that, if 
$\beta_L\neq\beta_R$, there exists no strongly continuous one-parameter
automorphism group of the spin algebra with respect to which the XY NESS 
is an equilibrium [KMS] state). The Lagrange multiplier approach of 
\cite[p.\hspace{0.4mm}168]{ARS} (see also 
\cite[p.\hspace{0.4mm}5186]{ARRS}), set up for a chain of finite 
length, directly defines a NESS as the ground state of an effective 
Hamiltonian which differs from the original Hamiltonian by the conserved 
macroscopic energy current observable. As formally discussed in 
\cite[p.\hspace{0.4mm}168]{O}, the effective Hamiltonian constructed in 
this way is substantially different from the one corresponding to the XY 
NESS in that it only contains a finite subfamily of the infinite family of 
all the charges present in the latter situation. Moreover, since the 
XY NESS consists of left movers and right movers carrying the inverse 
temperatures $\beta_R$ and $\beta_L$ of the right and left reservoirs, 
respectively (cf. \cite[p.\hspace{0.4mm}1171]{AP}), the exponential decay 
of the transversal spin-spin correlation function (cf. 
\cite[p.\hspace{0.4mm}10]{AB}) contrasts with its weak power law behavior 
in the Lagrange multiplier approach.
\er

For the sake of completeness, we display the proof of Theorem 
\ref{thm:ness-2} given in \cite[p.\hspace{0.4mm}3437]{A}.

\vspace{2mm}

\bprf
Since $\fA$ is generated by the identity $1$ and the elements $a(f)$ for 
all $f\in\fh$, since $|\omega_\rd(\tau_\lm^t(A))|\le\|A\|$ for all 
$t\in\R$ and all $A\in\fA$, and since $\omega_\rd$ satisfies the 
determinantal factorization property \eqref{qf}, it is enough to study, 
for all $n\in\N$, the large time limit of 
\ba
\omega_\rd(\tau_\lm^t(a^\ast(f_n)\ldots a^\ast(f_1)a(g_1)\ldots a(g_n)))
=\det(\Omega(t)),
\ea
where the map $\Omega:\R\to\Mat(n,\C)$ is defined, for all $t\in\R$ and 
all $i,j\in\{1,\ldots,n\}$, by
\ba
\label{Omega}
\Omega_{ij}(t)
:=(\e^{\i t h_\lm}g_i, s_\rd \e^{\i t h_\lm}f_j),
\ea
and $\Mat(n,\C)$ stands for the set of complex $n\times n$ matrices.
Moreover, since, due to Proposition \ref{prop:spctr} {\it (a)} from 
Appendix \ref{app:hlm}, the singular continuous spectrum of $h_\lm$ is 
empty, we have $1=1_{\rm ac}(h_\lm)+1_{\rm pp}(h_\lm)$. Thus, 
inserting $1$ between the propagator and the wave function on both sides 
of the scalar product in \eqref{Omega}, we can write, for all $t\in\R$ and
all $i,j\in\{1,\ldots,n\}$, that
\ba
\Omega_{ij}(t)
=\Omega_{ij}^{\rm aa}(t)
+\Omega_{ij}^{\rm ap}(t)
+\Omega_{ij}^{\rm pa}(t)
+\Omega_{ij}^{\rm pp}(t), 
\ea
where the maps $\Omega^{\rm aa},\Omega^{\rm ap},\Omega^{\rm pa}, 
\Omega^{\rm pp}:\R\to\Mat(n,\C)$ are defined, for all $t\in\R$ and all 
$i,j\in\{1,\ldots,n\}$, by
\ba
\label{Omaa}
\Omega_{ij}^{\rm aa}(t)
&:=(\e^{\i t h_\lm}1_{\rm ac}(h_\lm)g_i, s_\rd \e^{\i t h_\lm}1_{\rm ac}(h_\lm)f_j),\\
\label{Omap}
\Omega_{ij}^{\rm ap}(t)
&:=(\e^{\i t h_\lm}1_{\rm ac}(h_\lm)g_i, s_\rd \e^{\i t h_\lm}1_{\rm pp}(h_\lm)f_j),\\
\label{Ompa}
\Omega_{ij}^{\rm pa}(t)
&:=(\e^{\i t h_\lm}1_{\rm pp}(h_\lm)g_i, s_\rd \e^{\i t h_\lm}1_{\rm ac}(h_\lm)f_j),\\
\label{Ompp}
\Omega_{ij}^{\rm pp}(t)
&:=(\e^{\i t h_\lm}1_{\rm pp}(h_\lm)g_i, s_\rd \e^{\i t h_\lm}1_{\rm pp}(h_\lm)f_j).
\ea

{\it Term $\Omega_{ij}^{\rm aa}$}\\
Using that $[h_\rd, s_\rd]=0$ which follows from \eqref{whd} and
\eqref{wsd}, the large time limit of \eqref{Omaa} yields the wave
operator contribution to $s_\lm$ in \eqref{slm} since, for all 
$i,j\in\{1,\ldots, n\}$, we have
\ba
\lim_{t\to\infty}\Omega_{ij}^{\rm aa}(t)
&=\lim_{t\to\infty}(\e^{-\i t h_\rd}\e^{\i t h_\lm}1_{\rm ac}(h_\lm)g_i, 
s_\rd \e^{-\i t h_\rd} \e^{\i t h_\lm}1_{\rm ac}(h_\lm)f_j)\nonumber\\
&=(g_i, w^\ast(h_\rd,h_\lm)s_\rd w(h_\rd,h_\lm)f_j).
\ea

{\it Terms $\Omega_{ij}^{\rm ap}$ and $\Omega_{ij}^{\rm pa}$}\\
The large time limit of \eqref{Omap} behaves, for all 
$i,j\in\{1,\ldots, n\}$, like
\ba
\label{est-ap}
\lim_{t\to\infty}|\Omega_{ij}^{\rm ap}(t)|
&\le \lim_{t\to\infty} \|1_{\rm pp}(h_\lm) s_\rd \e^{\i t h_\lm} 
1_{\rm ac}(h_\lm) g_i\| \|f_j\|\nonumber\\
&=0,
\ea
where we used the Riemann-Lebesgue lemma and the fact from Proposition 
\ref{prop:spctr} {\it (c)} of Appendix \ref{app:hlm} that 
$1_{\rm pp}(h_\lm)\in\mL^0(\fh)$. The estimate from \eqref{est-ap}
also holds true for the term \eqref{Ompa}, of course.

{\it Term $\Omega_{ij}^{\rm pp}$}\\
Since, if $\lambda\neq 0$,  the subspace $\ran(1_{\rm pp}(h_\lm))$ is spanned by the normalized eigenfunction
$f_\lm\in\fh$ corresponding to the eigenvalue $e_\lm$ given in Proposition
\ref{prop:spctr} {\it (c)} of Appendix \ref{app:hlm}, we have 
$\e^{\i t h_\lm}1_{\rm pp}(h_\lm)=\e^{\i t e_\lm}1_{\rm pp}(h_\lm)$ for 
all $t\in\R$. This implies, for all $t\in\R$ and all 
$i,j\in\{1,\ldots, n\}$, that \eqref{Ompp} reads as
\ba
\label{contr-pp}
\Omega_{ij}^{\rm pp}(t)
:=(g_i, 1_{\rm pp}(h_\lm) s_\rd 1_{\rm pp}(h_\lm)f_j).
\ea
Finally, since the map $\det:\Mat(n,\C)\to\C$ is continuous, we arrive
at the conclusion.
\eprf

We next show that, as soon as the magnetic field strength is nonvanishing, 
the magnetic NESS breaks translation invariance. In order to do so, we 
switch to momentum space
\ba
\label{mspace}
\widehat\fh
:=L^2([-\pi,\pi];\tfrac{\rd k}{2\pi})
\ea
by means of the unitary Fourier transform $\ff:\fh\to\widehat\fh$ which is
defined, as usual, by $\ff f:=\sum_{x\in\Z}f(x)\e_x$ for all $f\in\fh$, and
for all $x\in\Z$, the plane wave function $\e_x:[-\pi,\pi]\to\C$ is given by 
$\e_x(k):=\e^{\i k x}$ for all $k\in [-\pi,\pi]$. 

In the following, we will also use the notation 
$\widehat a:=\ff a\ff^\ast\in\mL(\widehat\fh)$ for all $a\in\mL(\fh)$. 
Hence, in momentum space, the XY Hamiltonian $\widehat h$ acts through
multiplication by the dispersion relation function $\eps:[-\pi,\pi]\to\R$ 
defined, for all $k\in[-\pi,\pi]$, by
\ba
\label{eps}
\eps(k)
:=\cos(k).
\ea

The function introduced next will capture the effect of a nonvanishing
strength of the local external magnetic field. Its action will also be
particularly visible in the description of the heat flux in Theorem 
\ref{thm:ec} below. 

\begin{figure}
\centering
\includegraphics[width=55mm,height=40mm]{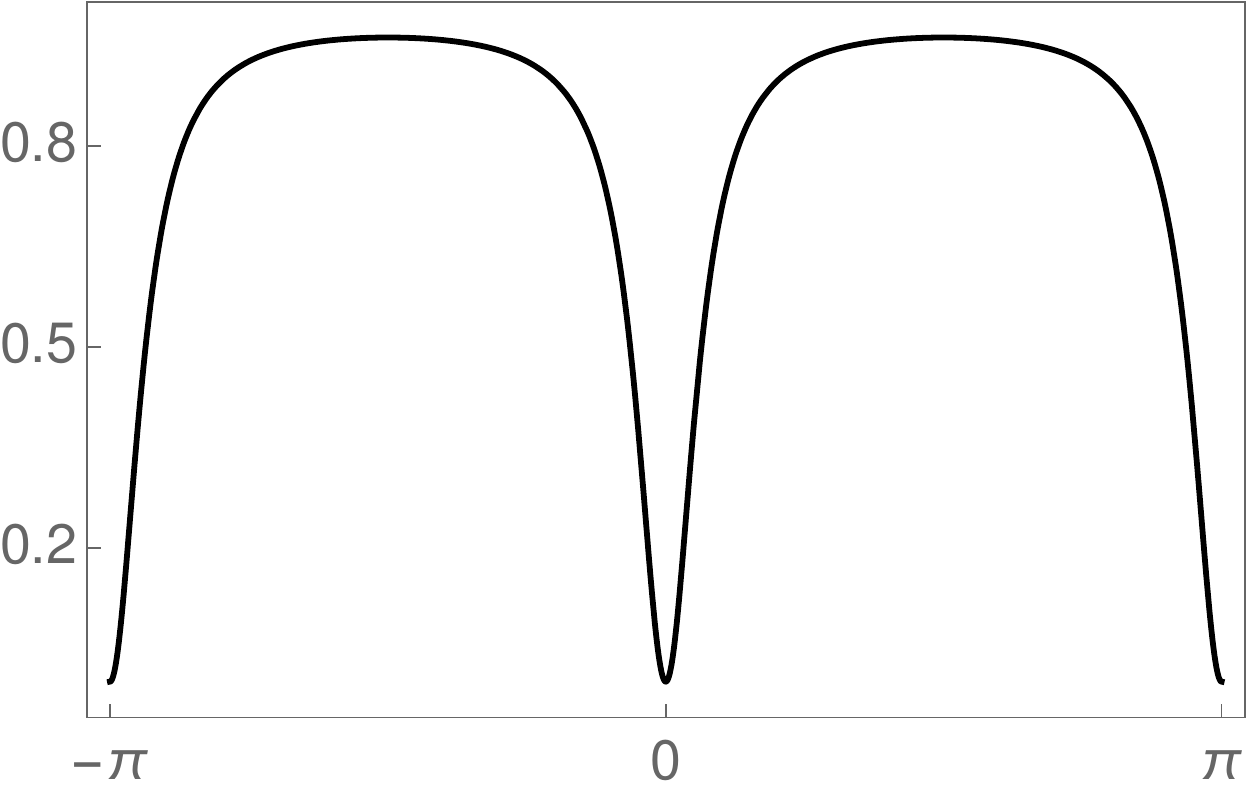}
\vspace{-2mm}
\caption{The magnetic correction function $[-\pi,\pi]\ni 
k\mapsto\Delta_\lm(\eps(k))\in\R$ for $\lm=\tfrac{1}{5}$.}
\label{fig:Dlm}
\end{figure}

\bd[Magnetic correction function]
The function $\Delta_\lm: [-1,1]\to\R$, defined, for all $e\in[-1,1]$, by
\ba
\label{DeltaB}
\Delta_\lm(e)
:=
\begin{cases}
\frac{1-e^2}{1-e^2+\lm^2}, & \lm\neq 0,\\
1, & \lm=0,
\end{cases}
\ea
is called the magnetic correction function (see Figure \ref{fig:Dlm}).
\ed

\br
The magnetic correction function also plays an important role in the 
mechanism which regularizes the symbol of the Toeplitz operator describing 
a certain class of nonequilibrium correlation functions in the magnetic 
NESS (see \cite[p.\hspace{0.4mm}3441]{A}).
\er

\bp[Broken translation invariance]
\label{prop:bti}
If $\lm\neq 0$, the magnetic NESS $\omega_\lm$ is breaking translation 
invariance.
\ep

\vspace{-5mm}

In order to carry out the proof, we will frequently refer to the results 
contained in the Appendices \ref{app:hlm}, \ref{app:xyness}, and 
\ref{app:w}.

\vspace{2mm}

\bprf
Due to Definition \ref{def:ti}, it is sufficient to show that 
$[s_\lm,u]\neq 0$ for all $\lm\in\R\setminus\{0\}$. To this end, we will
separately study the commutators of the absolutely continuous and the pure
point contributions to the two-point operator $s_\lm$ from \eqref{slm}.

{\it Term $1_{\rm ac}(h_\lm)$}\\
We want to take advantage of the fact used in 
\cite[p.\hspace{0.4mm}3439]{A} that the chain rule for wave operators 
allows us to write
\ba
\label{wsw}
w^\ast(h_\rd,h_\lm) s_\rd w(h_\rd,h_\lm)
&=w^\ast(h,h_\lm)w^\ast(h_\rd,h) s_\rd w(h_\rd,h)w(h,h_\lm)\nonumber\\
&=w^\ast(h,h_\lm) s w(h,h_\lm),
\ea
where $s$ is the so-called XY two-point operator, i.e., the two-point
operator inducing the XY NESS from Theorem \ref{thm:xyness} of Appendix 
\ref{app:xyness} and discussed in Remark \ref{rem:xyness}. Moreover, we 
also know from this theorem that $\widehat s$ acts in momentum space $\widehat\fh$ 
through multiplication by an explicit function $\theta$ given in 
\eqref{theta}. Hence, in order to compute the matrix elements of the commutator of \eqref{wsw} with $u$ with respect to the Kronecker basis 
$\{\delta_x\}_{x\in\Z}$ of $\fh$, we switch to momentum space and get, for
all $x, y\in\Z$, that
\ba
\label{wu}
(\delta_x,[w^\ast(h,h_\lm) s w(h,h_\lm),u]\delta_y)
&=(\widehat w(h,h_\lm)\e_x, \widehat s\hspace{0.4mm} \widehat w(h,h_\lm)\e_{y+1})\nonumber\\
&\hspace{5mm}-(\widehat w(h,h_\lm)\e_{x-1}, \widehat s\hspace{0.4mm}  \widehat w(h,h_\lm)\e_y).
\ea
Using the action of the wave operator $\widehat w(h,h_\lm)$ from \eqref{wm} in
Proposition \ref{prop:wm} of Appendix \ref{app:w} on the plane wave functions
$\e_x=\ff\delta_x$ for all $x\in\Z$, we find, for all $x, y\in\Z$, that
\ba
\label{wxy}
(\widehat w(h,h_\lm)\e_x,\widehat s\hspace{0.4mm}\widehat w(h,h_\lm)\e_y)
&=\int_{-\pi}^\pi\frac{\rd k}{2\pi}\,\theta(k)\,\e^{\i k(y-x)}\nonumber\\
&\hspace{4mm}+\i\lm \int_{-\pi}^\pi\frac{\rd k}{2\pi}\,\theta(k)\,
\frac{\sin(|k|)}{\sin^2(k)+\lm^2}\hspace{0.2mm}\left(\e^{\i(|ky|-kx)}-\e^{\i(ky-|kx|)}\right)\nonumber\\
&\hspace{4mm}-\lm^2\int_{-\pi}^\pi\frac{\rd k}{2\pi}\,\,
\frac{\theta(k)}{\sin^2(k)+\lm^2}\hspace{0.2mm}\left(\e^{\i(|ky|-kx)}+\e^{\i(ky-|kx|)}
-\e^{\i|k|(|y|-|x|)}\right).
\ea
We next plug $x=0$ and $y=1$ into \eqref{wu}. Separating the positive and negative momentum contributions in \eqref{wxy}, regrouping with respect 
to the temperature dependence of $\theta$, and reassembling the terms over
the whole momentum interval with the help of the evenness of $\eps$ from
\eqref{eps}, we find, for all $\lm\in\R\setminus\{0\}$, that
\ba
\label{ti2}
(\delta_0,[w^\ast(h,h_\lm) s w(h,h_\lm),u]\delta_1)
&=\i\lm  \int_{-\pi}^\pi\frac{\rd k}{2\pi}\,\, \theta(k)\,
\frac{\sin(|k|)}{\sin^2(k)+\lm^2}\,\left(\e^{2\i |k|}-\e^{2\i k}\right)
\nonumber\\
&\hspace{4mm}-\lm^2\int_{-\pi}^\pi\frac{\rd k}{2\pi}\,\,
\frac{\theta(k)}{\sin^2(k)+\lm^2}\hspace{1mm}\e^{2\i k}\nonumber\\
&\hspace{4mm}-\i\lm  \int_{-\pi}^\pi\frac{\rd k}{2\pi}\,\, \theta(k)\,
\frac{\sin(|k|)}{\sin^2(k)+\lm^2}\,\left(\e^{\i(|k|+k)}-\e^{\i(k-|k|)}
\right)\nonumber\\
&\hspace{4mm}+\lm^2\int_{-\pi}^\pi\frac{\rd k}{2\pi}\,\,
\frac{\theta(k)}{\sin^2(k)+\lm^2}\,\left(\e^{\i(|k|+k)}+\e^{\i(k-|k|)}-1
\right)\nonumber\\
&=\lm \int_{-\pi}^\pi\frac{\rd k}{2\pi}\,\, 
\eps(k)[\rho_{\beta_L}(\eps(k))-\rho_{\beta_R}(\eps(k))]\,
\Delta_\lm(\eps(k)).
\ea

{\it Term $1_{\rm pp}(h_\lm)$}\\
Due to Proposition \ref{prop:spctr} {\it (c)} of Appendix \ref{app:hlm},
we know that the pure point subspace of the magnetic Hamiltonian is 
spanned by the exponentially localized eigenfunction $f_\lm\in\fh$ given 
in \eqref{flm}. Hence, for all $x, y\in\Z$, we can write
\ba
\label{ppxy}
(\delta_x, [1_{\rm pp}(h_\lm)s_\rd 1_{\rm pp}(h_\lm),u]\delta_y)
=(f_\lm, s_\rd f_\lm) [f_\lm(x) f_\lm(y+1)-f_\lm(x-1) f_\lm(y)].
\ea 
Plugging $x=0$ and $y=1$ into \eqref{ppxy} and using the form of $f_\lm$ 
from \eqref{flm}, we get that 
\ba
\label{ti3}
(\delta_0, [1_{\rm pp}(h_\lm)s_\rd 1_{\rm pp}(h_\lm),u]\delta_1)
=0.
\ea
Hence, with the help of \eqref{ti2}, \eqref{DeltaB}, and \eqref{ti3}, we
arrive at the conclusion.
\eprf

\br
Since
$1/(1+\e^x)-1/(1+\e^y)=\sinh[(y-x)/2]/(\cosh[(y-x)/2]+\cosh[(y+x)/2)])$ 
for all $x,y\in\R$, the difference of the Planck density functions in
\eqref{ti2} reads, for all $x\in\R$, as
\ba
\label{dplanck}
\rho_{\beta_L}(x)-\rho_{\beta_R}(x)
=\frac{\sinh(\delta x)}{\cosh(\delta x)+\cosh(\beta x)},
\ea
where we used the definitions \eqref{delta} and \eqref{beta}.
\er

\section{Entropy production}
\label{sec:ep}

One of the central physical notions for systems out of equilibrium is the
well-known entropy production rate being a bilinear form in the affinities,
driving the system out of equilibrium, and in the fluxes, describing the
response to these applied forces. In the present situation, these 
quantities correspond to the differences between the inverse temperature 
of each reservoir and the inverse temperature of the sample and to the
corresponding heat fluxes (see, for example, \cite[p.\hspace{0.4mm}8]{R} 
and \cite[p.\hspace{0.4mm}1159]{AP}, and references therein). Hence, we 
are led to the following definition.

\bd[Energy current observable]
Let $\alpha\in\{L,R\}$. The one-particle energy current observable 
$\vi_\alpha\in\mL(\fh)$, describing the energy flow from reservoir $\alpha$
into the sample, is defined by
\ba
\label{def:opc}
\vi_\alpha
:=-\left.\frac{\rd}{\rd t}\right|_{t=0} 
\e^{\i t h_\lm} i_\alpha h_\alpha i_\alpha^\ast \e^{-\i t h_\lm}.
\ea
Moreover, the extensive energy current observable 
$\Phi_\alpha:=\rd\Gamma(\vi_\alpha)$ is the usual second 
quantization of $\vi_\alpha$.
\ed

The entropy production rate then reads as follows.

\bd[Entropy production rate]
The entropy production rate in the magnetic NESS $\omega_\lm$ is defined 
by
\ba
\label{ep0}
\sigma_\lm
:=-\sum_{\alpha\in\{L,R\}}\beta_\alpha\hspace{0.5mm}J_{\lm,\alpha},
\ea
where the NESS expectation value of the extensive energy current 
observable, the so-called heat flux, has been denoted, for all 
$\alpha\in\{L,R\}$, by
\ba
J_{\lm,\alpha}
:=\omega_\lm(\Phi_\alpha).
\ea
\ed

\br
Since, for all $\alpha\in\{L,R\}$,  we have 
$\vi_\alpha=-\i[h_\lm, i_\alpha h_\alpha i_\alpha^\ast]
=-\i[h, i_\alpha h_\alpha i_\alpha^\ast]$, we see that $\vi_\alpha$ is
independent of $\lm$ and that
\ba
\label{viL}
\vi^{}_L
&=\frac12\hspace{0.7mm}\Im(u^{-\nu} p_0 u^{\nu+2}),\\
\label{viR}
\vi^{}_R
&=\frac12\hspace{0.7mm}\Im(u^\nu p_0 u^{-(\nu+2)}),
\ea
i.e., $\vi_\alpha\in\mL^0(\fh)$ which implies that 
$\Phi_\alpha\in\fA(\fh)$ (the latter 
statement also holds in the more general case of trace class operators in 
the selfdual setting, see \cite[p.\hspace{0.5mm}410, 412]{Araki71}). Hence, 
$\sigma_\lm$ is well-defined.
\er

\br
Using that
$\sum_{\alpha\in\{L,R\}}i_\alpha^{} h_\alpha^{} i_\alpha^\ast
=h-i_S^{} h_S^{} i_S^\ast-(v_L+v_R)$, we can write
\ba
\sum_{\alpha\in\{L,R\}}\vi_\alpha
=\i [h_\lm,q],
\ea
where we set $q:=i_S^{} h_S^{} i_S^\ast+v_L+v_R+\lm v$. Hence, since 
$q\in\mL^0(\fh)$, we can write
\ba
\label{sumPhi}
\sum_{\alpha\in\{L,R\}}\Phi_\alpha
&=\rd\Gamma(\i [h_\lm, q])\nonumber\\
&=\frac{\rd}{\rd t}\bigg|_{t=0} 
\tau_\lm^t(\rd\Gamma(q)).
\ea
Taking the NESS expectation value of \eqref{sumPhi} and using that, due to
\eqref{mness}, $\omega_\lm$ is $\tau_\lm^t$-invariant for all $t\in\R$, 
we find the first law of thermodynamics, \ie, 
\ba
\label{fltd}
\sum_{\alpha\in\{L,R\}}\hspace{-1mm}J_{\lm,\alpha}
=0.
\ea
\er

Due to \eqref{ep0} and \eqref{fltd}, we restrict ourselves to the study of
the NESS expectation value of the extensive energy current describing the
energy flow from the left reservoir into the sample, and we will use the simplified notation $J_\lm:=J_{\lm,L}$ in the following.

\begin{figure}
\centering
\includegraphics[width=55mm,height=40mm]{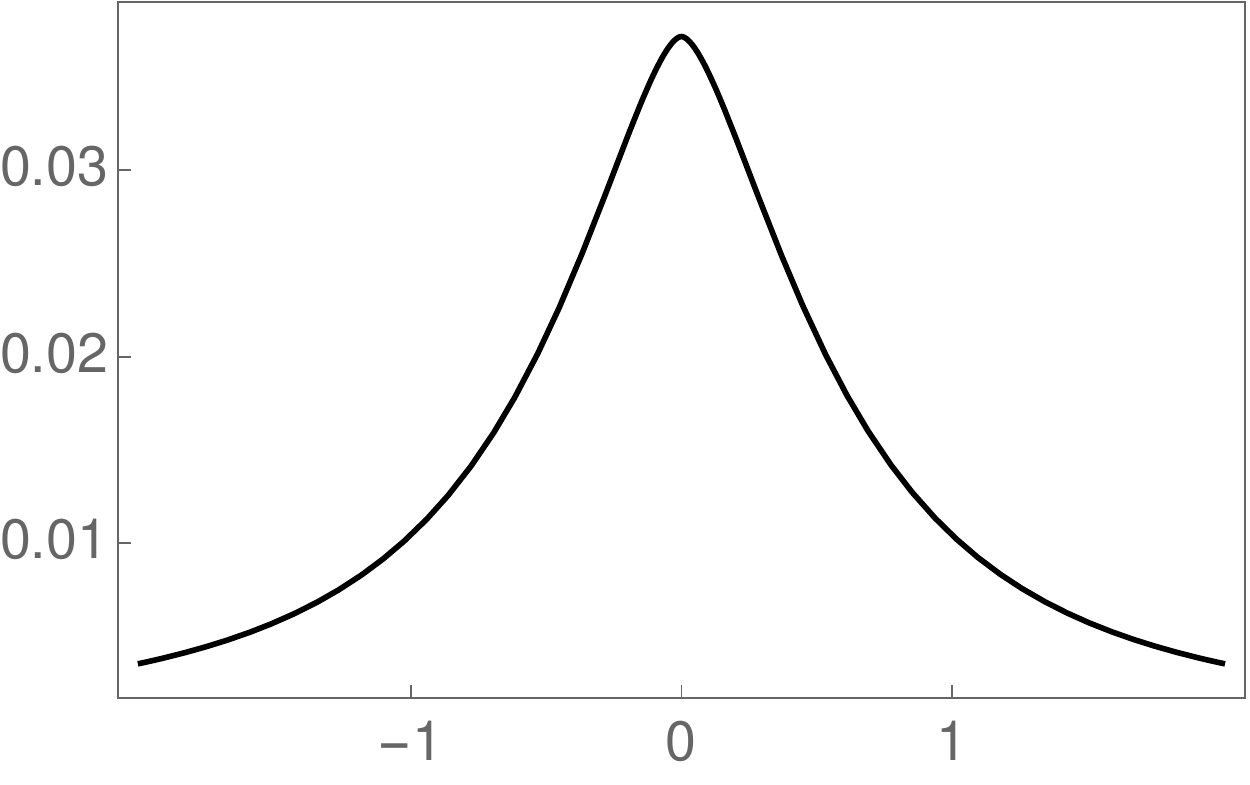}
\vspace{-2mm}
\caption{The heat flux $\R\ni\lm\mapsto J_\lm\in\R$ as a function of the
strength of the local external magnetic field for $\beta_L=1$ and 
$\beta_R=2$.}
\label{fig:J}
\end{figure}

\bt[Heat flux]
\label{thm:ec}
The heat flux, \ie, the NESS expectation value of the extensive energy current observable describing the energy flow from the left reservoir 
into the sample has the form
\ba
\label{J}
J_\lm
=\frac12\int_{-\pi}^\pi\frac{\rd k}{2\pi}\hspace{2mm}
j(\eps(k))\, \Delta_\lm(\eps(k)),
\ea
where the density function $j:[-1,1]\to\R$ is defined, for all 
$e\in [-1,1]$, by
\ba
\label{j}
j(e)
:=e\hspace{0.2mm}\sqrt{1-e^2}\hspace{0.3mm}
\big[\rho_{\beta_L}(e)-\rho_{\beta_R}(e)\big],
\ea
and the functions $\rho_{\beta_L}, \rho_{\beta_R}$, $\eps$, and 
$\Delta_\lm$ are given in \eqref{rhor}, \eqref{eps}, and \eqref{DeltaB},
respectively (see Figure \ref{fig:J}).
\et

\br
Since, due to the proof of Theorem \ref{thm:nqpt} {\it (a)} below, we have
that $(j\Delta_\lm)\circ\eps\in L^1([-\pi,\pi];\rd k)$, the integral 
\eqref{J} is well-defined.
\er

\br
Note that \eqref{J} is independent of the sample size parameter $\nu$.
\er

\bprf
The case $\lm=0$ is treated in \cite[p.\hspace{0.4mm}1160]{AP} (see Remark
\ref{rem:xy} above) and corresponds to the triviality of the magnetic correction function $\Delta_0=1$ in \eqref{J}.

In the following, we thus discuss the case $\lm\neq 0$. Using \eqref{slm} 
and the fact that $\omega_\lm(\rd\Gamma(\vi))=\tr(s_\lm\vi)$ for all 
$\vi\in\mL^0(\fh)$ (a similar statement also holds in the more general case of
trace class operators in the selfdual setting, see 
\cite[p.\hspace{0.5mm}410, 412]{Araki71}), the NESS expectation value of 
the extensive energy current observable describing the energy flow from 
the left reservoir into the sample can be written in the form
\ba
\label{JL}
J_\lm
&=\omega_\lm(\Phi_L)\nonumber\\
&=\omega_\lm(\rd\Gamma(\vi_L))\nonumber\\
&=\tr(s_\lm\vi_L)\nonumber\\
&=J_{\lm, \rm ac}+J_{\lm, \rm pp},
\ea
where we make the definitions
\ba
\label{Jac0}
J_{\lm, \rm ac}
&:=\tr(w^\ast(h_\rd,h_\lm) s_\rd w(h_\rd,h_\lm)\vi_L),\\
\label{Jpp0}
J_{\lm, \rm pp}
&:=\tr(1_{\rm pp}(h_\lm) s_\rd 1_{\rm pp}(h_\lm)\vi_L).
\ea
We next discuss the two terms \eqref{Jac0} and  \eqref{Jpp0} separately.

{\it Term $J_{\lm, \rm ac}$}\\
Using \eqref{wsw}, \eqref{viL}, the Kronecker basis 
$\{\delta_x\}_{x\in\Z}$ of $\fh$ and the fact that the two-point operator 
$s$ is selfadjoint, we can write
\ba
\label{Jcomp}
J_{\lm, \rm ac}
&=\frac12\,\tr(w^\ast(h,h_\lm)s w(h,h_\lm)\Im(u^{-\nu} p_0 u^{\nu+2}))
\nonumber\\
&=\frac{1}{2}\,\Im[(w(h,h_\lm)\delta_{-(\nu+2)},s w(h,h_\lm)\delta_{-\nu})].
\ea
In order to compute \eqref{Jcomp}, we proceed as in the proof of 
Proposition \ref{prop:bti} and switch to momentum space $\widehat\fh$. 
Using \eqref{wxy} for $x=-(\nu+2)$ and $y=-\nu$, we then get
\ba
(\widehat w(h,h_\lm)\e_{-(\nu+2)},\widehat s\hspace{0.5mm}\widehat 
w(h,h_\lm)\e_{-\nu})
=\sum_{i=1}^3 z_i,
\ea
where we make the definitions
\ba
\label{z1}
z_1
&:=\int_{-\pi}^\pi\frac{\rd k}{2\pi}\,\theta(k)\,\e^{2\i k}\\
\label{z2}
z_2
&:=\i\lm \int_{-\pi}^\pi\frac{\rd k}{2\pi}\,\theta(k)\,
\frac{\sin(|k|)}{\sin^2(k)+\lm^2}\left(\e^{\i(|k|\nu+k(\nu+2))}
-\e^{-\i(k\nu+|k|(\nu+2))}\right)\\
\label{z3}
z_3
&:=-\lm^2\int_{-\pi}^\pi\frac{\rd k}{2\pi}\,
\frac{\theta(k)}{\sin^2(k)+\lm^2}
\left(\e^{\i(|k|\nu+k(\nu+2))}+\e^{-\i(k\nu+|k|(\nu+2))}-\e^{-2\i|k|}\right),
\ea
and we recall that $\theta$ is given in \eqref{theta} of Theorem 
\ref{thm:xyness} in Appendix \ref{app:xyness}. Separating the positive and 
negative momentum contributions (which directly leads to \eqref{Imz2}),
regrouping with respect to the temperature dependence of $\theta$, and
reassembling the terms over the whole momentum interval with the help of 
the evenness of $\eps$, the imaginary parts take the form
\ba
\label{Imz1}
\Im(z_1)
&=\int_{-\pi}^\pi\frac{\rd k}{2\pi}\,\cos(k) |\sin(k)|
[\rho_{\beta_L}(\cos(k))-\rho_{\beta_R}(\cos(k))],\\
\label{Imz2}
\Im(z_2)
&=0,\\
\label{Imz3}
\Im(z_3)
&=-\lm^2\int_{-\pi}^\pi\frac{\rd k}{2\pi}\,\cos(k) |\sin(k)|
[\rho_{\beta_L}(\cos(k))-\rho_{\beta_R}(\cos(k))]\,\frac{1}{\sin^2(k)+\lm^2}.
\ea
Therefore, \eqref{Imz1}, \eqref{Imz2}, and \eqref{Imz3} lead us to 
\ba
\label{Jac}
J_{\lm, \rm ac}
=\frac12\int_{-\pi}^\pi\frac{\rd k}{2\pi}\,\cos(k) |\sin(k)|
[\rho_{\beta_L}(\cos(k))-\rho_{\beta_R}(\cos(k))]\,\frac{\sin^2(k)}{\sin^2(k)+\lm^2}.
\ea

{\it Term $J_{\lm, \rm pp}$}\\
Since the one-particle energy current observable 
$\vi_L=-\i[h_\lm, i_L^{}h_L^{}i_L^\ast]$ has the form of a commutator, the
cyclicity of the trace implies that 
\ba
\label{Jpp}
J_{\lm,\rm pp}
&=\tr(s_\rd1_{\rm pp}(h_\lm)\vi_L 1_{\rm pp}(h_\lm))\nonumber\\
&=-\i\,\tr(s_\rd1_{\rm pp}(h_\lm)[h_\lm, i_L^{}h_L^{}i_L^\ast]  1_{\rm pp}(h_\lm))\nonumber\\
&=-\i\, \tr(s_\rd1_{\rm pp}(h_\lm)[e_\lm 1, i_L^{}h_L^{}i_L^\ast]  1_{\rm pp}(h_\lm))\nonumber\\
&=0,
\ea
where $e_\lm$ is the unique eigenvalue of the magnetic Hamiltonian 
$h_\lm$ given in Proposition \ref{prop:spctr} {\it (c)} of Appendix 
\ref{app:hlm}. 

Hence, using \eqref{Jac}, \eqref{Jpp}, \eqref{DeltaB}, and \eqref{j}, we
arrive at the conclusion.
\eprf

\bc[Strict positivity of the entropy production]
The heat flux is flowing through the sample from the hotter to the
colder reservoir. Moreover, the entropy production rate is strictly 
positive, 
\ba
\sigma_\lm>0.
\ea
\ec

\bprf
It immediately follows from \eqref{J} and the form of the functions 
\eqref{DeltaB} and \eqref{j} that $J_\lm>0$. Moreover, using \eqref{ep0} 
and \eqref{fltd}, the entropy production can be written
as
\ba
\label{ep}
\sigma_\lm
=(\beta_R-\beta_L) J_\lm.
\ea
Hence, we 
arrive at the conclusion.
\eprf

\section{Nonequilibrium quantum phase transition}
\label{sec:nqpt}

Going beyond Ehrenfest's classification proposition not only with respect to
the nature of the singularity, as it is usually done in the modern 
classification schemes, but also with respect to the nature of the 
state considered, we define a nonequilibrium quantum phase transition to 
be a point of (higher-order) non-differentiability of the entropy production 
rate with
respect to an external physical parameter of interest. In the case at 
hand, the parameter which we are interested in is the local external 
magnetic field $\lm$. 

In the following, $J'_\lm$ will denote the derivative of $J_\lm$ with
respect to $\lm$.

\bt[Second-order nonequilibrium quantum phase transition]
\label{thm:nqpt}
The heat flux $\R\ni\lm\mapsto J_\lm\in\R$ has the following properties:
\bn
\ita
It belongs to $C^1(\R)\cap C^\infty(\R\setminus\{0\})$.

\itb
Its second derivative with respect to the local external magnetic field 
does not exist at the origin, \ie, for $\lm\to 0$, we have
\ba
J_\lm'
=C\lm\log(|\lm|)+\mO(\lm),
\ea
where we set $C:=(4/\pi)[\rho(\beta_L)-\rho(\beta_R)]$.
\en
\et

\vspace{3mm}

\br
In \cite[p.\hspace{0.4mm}4]{ABZ}, an interesting study of a class of
nonequilibrium quantum phase transitions in the NESS constructed in 
\cite[p.\hspace{0.4mm}1158]{AP} has been carried out. In particular,
it has been shown that current type correlation functions in this 
NESS have a discontinuous first and third derivative with respect to the
spatially homogeneous external magnetic field $\mu$ (at different 
$(\gamma,\mu)$-values on the so-called nonequilibrium critical line) 
as soon as the anisotropy $\gamma$ is nonvanishing (cf. Remark 
\ref{rem:xy}).
\er

\br
In the Lagrange multiplier approach discussed in Remark 
\ref{rem:approaches}, the expectation value of the macroscopic energy 
current observable in the ground state of the effective Hamiltonian 
defines a phase diagram in the effective field-magnetic field plane
(cf. \cite[p.\hspace{0.4mm}169]{ARS}) separating the equilibrium phase,
in which the heat flux is zero, from the nonequilibrium phase of 
nonvanishing heat flux (in the latter phase, the spin-spin correlations 
in the $x$ direction also exhibit an oscillatory behavior dominated by
a slowly decaying amplitude). This interpretation contrasts with Theorem
\ref{thm:nqpt} which yields a point of second-order non-differentiability 
of the always strictly positive entropy production rate.
\er

\vspace{1mm}

\bprf
We start off by noting that the function $f:\R\times [-\pi,\pi]\to\R$, 
defined, for all $\lm\in\R$ and all $k\in [-\pi,\pi]$, by
\ba
\label{f}
f(\lm,k)
&:=j(\eps(k))\, \Delta_\lm(\eps(k))\nonumber\\
&=j(\cos(k))\cdot 
\begin{cases}
\frac{\sin^2(k)}{\sin^2(k)+\lm^2}, & \lm\neq 0,\\
1, & \lm=0,
\end{cases}
\ea
is integrable in $k$ over $[-\pi,\pi]$ for any fixed $\lm\in\R$ since 
$|j(\cos(k))|\le 2$ for all $k\in[-\pi,\pi]$ and since
$\sin^2(k)/(\sin^2(k)+\lm^2)\le 1$ for all $\lm\neq 0$ and all 
$k\in[-\pi,\pi]$.

Using Lebesgue's dominated convergence theorem, we can now proceed to
study the regularity of the heat flux.

\bn
\ita
{\it $J_\cdot\in C(\R)$}\\
We first want to show that $f(\,\cdot\hspace{0.5mm},k)\in C(\R)$ for any 
fixed $k\in [-\pi,\pi]$. In order to do so, 
we note that, since 
$j(\cos(k))=0$ and $\Delta_\lm(\cos(k))=\delta_{0\lm}$ for all 
$k\in\{0,\pm\pi\}$ and all $\lm\in\R$, we get that 
$f(\,\cdot\hspace{0.5mm},k)=0$ for all $k\in\{0,\pm\pi\}$, \ie, 
$f(\,\cdot\hspace{0.5mm},k)\in C(\R)$ for all $k\in\{0,\pm\pi\}$. 
Moreover, \eqref{f} immediately implies that 
$f(\,\cdot\hspace{0.5mm},k)\in C(\R)$ for all 
$k\in[-\pi,\pi]\setminus\{0,\pm\pi\}$. Second, since we have from above
that $|f(\lm,k)|\le 2\in L^1([-\pi,\pi];\rd k)$ for all $\lm\in\R$ and all 
$k\in[-\pi,\pi]$, Lebesgue's dominated convergence theorem implies that
$J_\cdot\in C(\R)$.

{\it $J_\cdot\in C^1(\R)$}\\
Again, first, we immediately get that $f(\,\cdot\hspace{0.5mm},k)\in C^1(\R)$ 
for all $k\in\{0,\pm\pi\}$ and 
$f(\,\cdot\hspace{0.5mm},k)\in C^1(\R\setminus\{0\})$ 
for all $k\in[-\pi,\pi]\setminus\{0,\pm\pi\}$. Moreover, for any fixed
$k\in[-\pi,\pi]\setminus\{0,\pm\pi\}$, we have 
$\lim_{\lm\to 0}(f(\lm,k)-f(0,k))/\lm=0$. Therefore, for all 
$k\in[-\pi,\pi]\setminus\{0,\pm\pi\}$ and all $\lm\in\R$, we get
\ba
\label{df}
\frac{\partial f}{\partial\lm}(\lm,k)
=-2\lm j(\cos(k))\,\frac{\sin^2(k)}{(\sin^2(k)+\lm^2)^2},
\ea
which implies that $f(\,\cdot\hspace{0.5mm},k)\in C^1(\R)$ 
for all $k\in[-\pi,\pi]\setminus\{0,\pm\pi\}$. Second, since 
$0\le(|\sin(k)|-|\lm|)^2=\sin^2(k)-2|\lm||\sin(k)|+\lm^2$ for all 
$\lm\in\R$ and all $k\in [-\pi,\pi]$, we can write 
$2|\lm||\sin(k)|^3\le (\sin^2(k)+\lm^2)^2$ for all 
$\lm\in\R$ and all $k\in [-\pi,\pi]$. Hence, since 
$|j(\cos(k))|\le 2|\cos(k)||\sin(k)|$ for all $k\in [-\pi,\pi]$, it 
follows, for all $k\in[-\pi,\pi]\setminus\{0,\pm\pi\}$ and all $\lm\in\R$,
that 
\ba
\left|\frac{\partial f}{\partial\lm}(\lm,k)\right|
&\le 2\,|\!\cos(k)|\, \frac{2|\lm||\sin(k)|^3}{ (\sin^2(k)+\lm^2)^2},
\ea
which leads to 
$|\tfrac{\partial f}{\partial\lm}(\lm,k)|\le 2\in L^1([-\pi,\pi];\rd k)$ 
for all $\lm\in\R$ and all $k\in[-\pi,\pi]$. Hence, Lebesgue's dominated 
convergence theorem implies that $J_\cdot\in C^1(\R)$.

{\it $J_\cdot\in C^\infty(\R\setminus\{0\})$}\\
Since \eqref{f} tells us that, for all $k\in [-\pi,\pi]$, the derivatives 
of any order of $f(\,\cdot\hspace{0.5mm},k)$ with respect to $\lm$ exist 
and are bounded by a constant for all $k\in [-\pi,\pi]$ and all points in 
a sufficiently small neighborhood of any $\lm\neq 0$, Lebesgue's dominated
convergence theorem also yields that 
$J_\cdot\in C^\infty(\R\setminus\{0\})$.

\begin{figure}
\centering
\includegraphics[width=55mm,height=40mm]{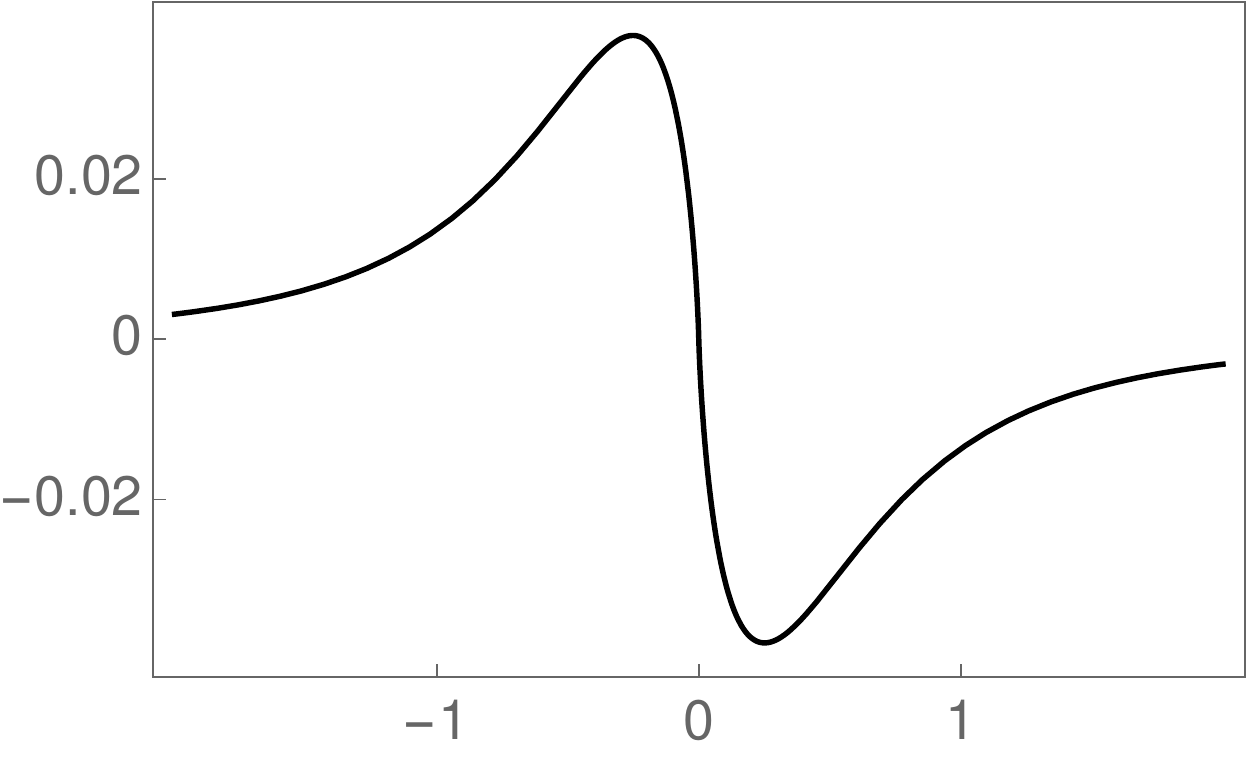}
\vspace{-2mm}
\caption{The first derivative of the heat flux with respect to the 
strength of the local external magnetic field  
$\R\ni\lm\mapsto J_\lm'\in\R$ for 
$\beta_L=1$ and $\beta_R=2$.}
\label{fig:Jp}
\end{figure}

\begin{figure}
\centering
\includegraphics[width=55mm,height=40mm]{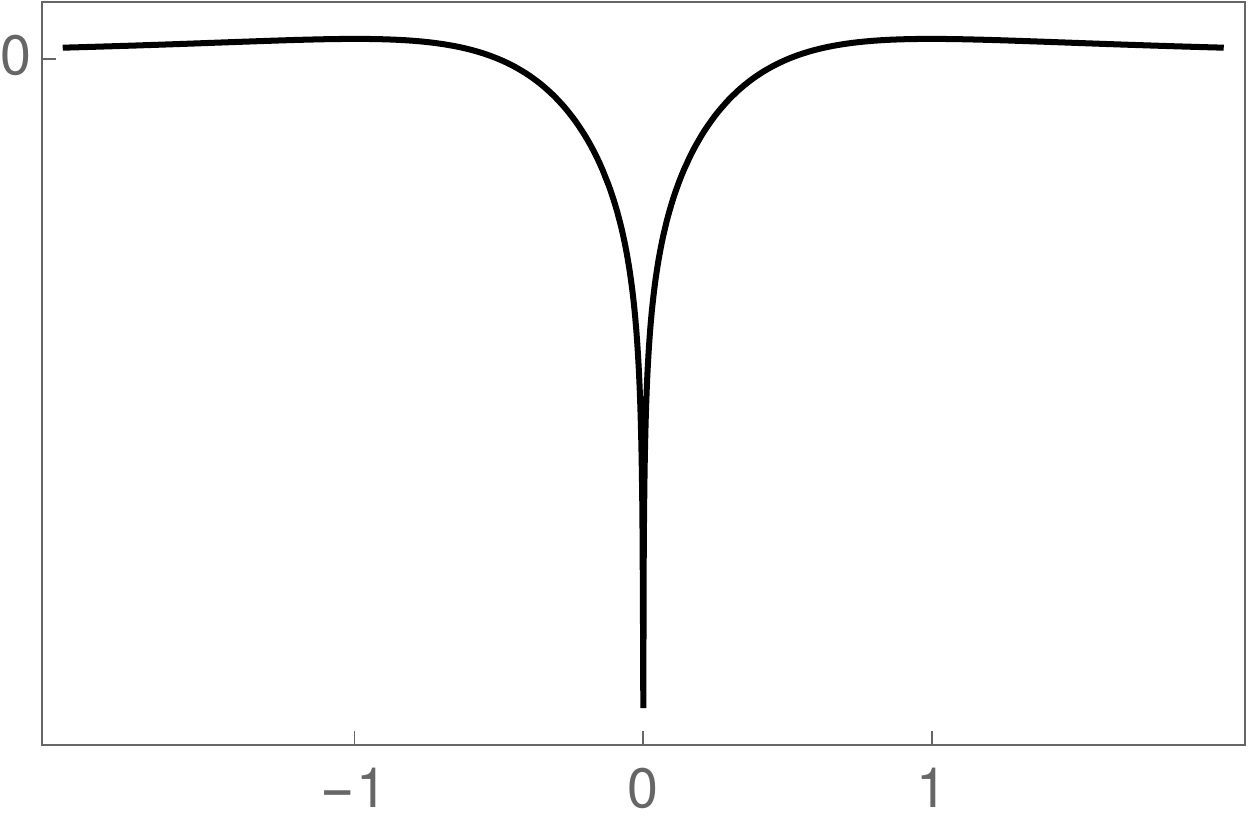}
\vspace{-2mm}
\caption{The second derivative of the heat flux with respect to the 
strength of the local external magnetic field $\R\setminus\{0\}\ni\lm\mapsto J_\lm''\in\R$ for $\beta_L=1$ and $\beta_R=2$.}
\label{fig:Jpp}
\end{figure}

\itb
Using {\it (a)},  \eqref{df}, \eqref{j}, and \eqref{dplanck}, we get, for 
all $\lm\in\R$, that
\ba
\label{sp}
J_\lm'
&=-2\lm \int_{-\pi}^\pi\frac{\rd k}{2\pi}\hspace{2mm}
\cos(k) |\sin(k)|[\rho_{\beta_L}(\cos(k))-\rho_{\beta_R}(\cos(k))]
\hspace{0.5mm}
\frac{\sin^2(k)}{(\sin^2(k)+\lm^2)^2}\nonumber\\
&=-8\lm \int_0^{\frac\pi2}\frac{\rd k}{2\pi}\hspace{2mm}
\frac{\cos(k)\, \sinh(\delta \cos(k))}
{\cosh(\delta \cos(k))+\cosh(\beta\cos(k))}
\hspace{1mm}
\frac{\sin^3(k)}{(\sin^2(k)+\lm^2)^2},
\ea
where, in the second equality, we reduced the integration interval from 
$[-\pi,\pi]$ to $[0,\pi]$ and then from $[0,\pi]$ to $[0,\pi/2]$ by
using the coordinate transformations $k\mapsto -k$ and $k\mapsto \pi-k$,
respectively (see Figure \ref{fig:Jp}). Next, using the coordinate transformation 
$\arcsin:[0,1]\to [0,\tfrac\pi2]$ in \eqref{sp}, we can write, for all
$\lm\in\R\setminus\{0\}$, that (see Figure \ref{fig:Jpp})
\ba
\label{int1}
-\frac{\pi}{4}\frac{J_\lm'}{\lm}
=\int_0^1\rd x\,\,
f(x)\,\frac{x^3}{(x^2+\lm^2)^2},
\ea
where the function $f:[0,1]\to\R$ is defined, for all $x\in[0,1]$, by
\ba
f(x)
:=\frac{\sinh(\delta\sqrt{1-x^2})}{\cosh(\delta\sqrt{1-x^2})
+\cosh(\beta\sqrt{1-x^2})}.
\ea

In order to extract the logarithmic divergence at the origin, we make the
decomposition
\ba
\int_0^1\rd x\,\,
f(x)\,\frac{x^3}{(x^2+\lm^2)^2}
=F_1(\lm)+F_2(\lm),
\ea
where the functions $F_1, F_2:\R\setminus\{0\}\to\R$ are defined, for all
$\lm\in\R\setminus\{0\}$, by
\ba
\label{int2}
F_1(\lm)
&:=f(0) \int_0^1\rd x\,\,\frac{x^3}{(x^2+\lm^2)^2},\\
\label{int3}
F_2(\lm)
&:=\int_0^1\rd x\,\,[f(x)-f(0)]\,\frac{x^3}{(x^2+\lm^2)^2}.
\ea
{\it Term $F_1(\lm)$}\\ 
Using the primitive of its integrand, the first integral reads, for
all $\lm\in\R\setminus\{0\}$, as
\ba
\label{F1}
F_1(\lm)
&=-\frac{\sinh(\delta)}{\cosh(\delta)+\cosh(\beta)}\log(|\lm|)\nonumber\\
&\hspace{4.5mm}-\frac12\frac{\sinh(\delta)}{\cosh(\delta)+\cosh(\beta)}
\left(\frac{1}{1+\lm^2}-\log(1+\lm^2)\right),
\ea
and the second term on the right hand side of \eqref{F1} is obviously
defined and bounded in any neighborhood of the origin.

{\it Term $F_2(\lm)$}\\ 
In order to treat the term \eqref{int3}, we use that
$|f(x)-f(0)|\le \int_0^x\rd t\,\,|f'(t)|$ for all $x\in[0,1]$ and that the
derivative of $f$ can be estimated, for all $t\in[0,1)$, as
\ba
\label{est1}
|f'(t)|
&=\frac{ct}{\sqrt{1-t^2}},
\ea
where we set 
$c:=[\delta+\delta\cosh(\delta)\cosh(\beta)
+\beta\sinh(\delta)\sinh(\beta)]/4$. It then follows from 
\eqref{est1} that $|f(x)-f(0)|\le c x^2/(1+\sqrt{1-x^2})$ for all 
$x\in[0,1]$. Hence, \eqref{int3} can be bounded, for all 
$\lm\in\R\setminus\{0\}$, by
\ba
|F_2(\lm)|
&\le c\int_0^1\rd x\,\frac{x^4}{(x^2+\lm^2)^2}\nonumber\\
&\le c.
\ea
Finally, using the same estimates, Lebesgue's dominated convergence 
theorem also implies that $F_2$ has a continuous extension to the origin, 
the latter being again bounded by $c$ in any neighborhood of the origin. 
\en
Hence, we arrive at the conclusion.
\eprf

\begin{appendix}
\label{sec:app}
\section{Magnetic Hamiltonian}
\label{app:hlm}

In this appendix, we summarize the spectral theory of $h_\lm\in\mL(\fh)$ 
needed in the previous sections. To this end, we recall that 
$\spec_{\rm sc}(h_\lm)$, $\spec_{\rm ac}(h_\lm)$, and 
$\spec_{\rm pp}(h_\lm)$ stand for the singular continuous, the absolutely
continuous, and the point spectrum of $h_\lm$, respectively.

\bp[Magnetic spectrum]
\label{prop:spctr}
The magnetic Hamiltonian $h_\lm\in\mL(\fh)$ has the following spectral
properties:
\bn
\item[(a)] $\spec_{\rm sc}(h_\lm)=\emptyset$

\item[(b)] $\spec_{\rm ac}(h_\lm)=[-1,1]$

\item[(c)]
\label{sp-pp}
$\spec_{\rm pp}(h_\lm)
=\begin{cases}
\emptyset, & \lm=0\\
\{e_\lm\}, & \lm\neq0
\end{cases}$

Here, the eigenvalue is given by
\ba
e_\lm
:=\begin{cases}
\sqrt{1+\lm^2}, & \lm>0,\\
-\sqrt{1+\lm^2}, & \lm<0.
\end{cases}
\ea
Moreover, we have $\ran(1_{\rm pp}(h_\lm))={\rm span}\{f_\lm\}$, and
the normalized eigenfunction $f_\lm\in\fh$ is defined, for all $x\in\Z$,
by
\ba
\label{flm}
f_\lm(x)
:=\frac{\e^{-\alpha_\lm |x|}}{\nu_\lm}\cdot
\begin{cases}
1, & \lm>0,\\
(-1)^x, & \lm<0,
\end{cases}
\ea
where the inverse decay rate and the square of the normalization constant 
are given by $\alpha_\lm:=\log(\sqrt{1+\lm^2}+|\lm|)$ and 
$\nu_\lm^2:=\sqrt{1+\lm^2}/|\lm|$, respectively.
\en
\ep

\bprf
See \cite[p.\hspace{0.4mm}3449]{A} for the case $\lm>0$. An analogous
straightforward calculation also yields the eigenvalue and the 
eigenfunction for the case $\lm<0$ (the case $\lm=0$ corresponds to the
Laplacian on the discrete line).
\eprf
\section{XY NESS}
\label{app:xyness}

In \cite{AP}, we constructed the unique translation invariant NESS, called 
the XY NESS, in the sense of \eqref{mness} for the general XY model 
briefly discussed in Remark \ref{rem:xy}. For simplicity, we restrict the formulation of the following 
assertion to the case at hand for which the XY NESS with $\gamma=\mu=0$
corresponds to the limit \eqref{mness} with $\lm=0$. 

\bt[XY two-point operator]
\label{thm:xyness}
The XY NESS is the quasifree state induced by the two-point operator
$s=w^\ast(h_\rd,h) s_\rd w(h_\rd,h)$. In momentum space, $\widehat s$ acts 
through multiplication by the function $\theta:[-\pi,\pi]\to\C$ given  by
\ba
\label{theta}
\theta(k)
:=
\begin{cases}
\rho_{\beta_R}(\eps(k)), &k\in [-\pi,0],\\
\rho_{\beta_L}(\eps(k)), &k\in (0,\pi],
\end{cases}
\ea
and we recall that $\eps(k)=\cos(k)$ for all $k\in [-\pi,\pi]$.
\et

\bprf
See \cite[p.\hspace{0.4mm}1171]{AP}.
\eprf

\section{Scattering theory}
\label{app:w}

In \cite{A}, we determined the action in momentum space of the wave
operator \eqref{wave} on the completely localized orthonormal Kronecker
basis $\{\delta_x\}_{x\in\Z}$ of $\fh$ using the stationary scheme of
scattering theory and the weak abelian form of the wave operator. 
Recall that the plane wave function is related to the Kronecker 
function by $e_x=\ff\delta_x$ for all $x\in\Z$, where we used the
notations introduced after \eqref{mspace}.

\bp[Wave operator]
\label{prop:wm}
In momentum space, the action  of the wave operator\linebreak
$w(h,h_\lm)$ on the elements of the completely localized Kronecker basis 
$\{\delta_x\}_{x\in\Z}$ reads, for all $x\in\Z$ and all $k\in [-\pi,\pi]$, 
as
\ba
\label{wm}
({\widehat w}(h,h_\lm)\e_x)(k)
=\e_x(k)+\i\lm\hspace{1mm}\frac{\e_{|x|}(|k|)}{\sin(|k|)-\i\lm}.
\ea
\ep

\bprf
See \cite[p.\hspace{0.4mm}3439]{A}.
\eprf

\br
The action \eqref{wm} relates to the action of the wave operator for the
one-center $\delta$-interaction on the continuous line by replacing
$\sin(|k|)$ by $|k|$.
\er

\end{appendix}


\begin{thebibliography}{99}

\bibitem{ABZ} Ajisaka S, Barra F, and \v{Z}unkovi\v{c} B 2014
{\it Nonequilibrium quantum phase transitions in the XY model: comparison 
of unitary time evolution and reduced density matrix approaches}
New J. Phys. 16 033028 

\bibitem{ARRS} Antal T, R\'acz Z, R\'akos A, and Sch\"utz G M L 1998
{\it Isotropic transverse XY chain with energy and magnetization
currents}
Phys. Rev. E 57 5184

\bibitem{ARS} Antal T, R\'acz Z, and Sasv\'ari L 1997
{\it Nonequilibrium steady state in a quantum system: one-dimensional
transverse Ising model with energy current}
Phys. Rev. Lett. 78 167

\bibitem{Araki71} Araki H 1971 
{\it On quasifree states of CAR and Bogoliubov automorphisms} 
Publ. RIMS Kyoto Univ. 6 385

\bibitem{Araki84} Araki H 1984 
{\it On the XY-model on two-sided infinite chain}
Publ. RIMS Kyoto Univ. 20 277

\bibitem{AH} Araki H and Ho T G 2000 
{\it Asymptotic time evolution of a partitioned infinite two-sided isotropic XY-chain}
Proc. Steklov Inst. Math. 228 191  

\bibitem{A} Aschbacher W H 2011
{\it Broken translation invariance in quasifree fermionic correlations
out of equilibrium}
J. Funct. Anal. 260 3429

\bibitem{AB} Aschbacher W H and Barbaroux J-M 2007
{\it Exponential spatial decay of spin-spin correlations
in translation invariant quasifree states}
J. Math. Phys. 48 113302

\bibitem{AP} Aschbacher W H and Pillet C A 2003 
{\it  Non-equilibrium steady states of the XY chain} 
J. Stat. Phys. 112 1153

\bibitem{BR} Bratteli O and Robinson D W 1987/1997
{\it Operator algebras and quantum statistical mechanics 1/2}
(Springer)

\bibitem{HBC} Callen H B 1960 
{\it Thermodynamics}
(Wiley)

\bibitem{CSP} Culvahouse J W, Schinke D P, and Pfortmiller L G 1969
{\it Spin-spin interaction constants from the hyperfine structure of coupled 
ions}
Phys. Rev. 177 454 

\bibitem{LSM}Lieb E, Schultz T, and Mattis D 1961 
{\it Two soluble models of an antiferromagnetic chain}
Ann. Physics 16 407 

\bibitem{MO} Matsui T and Ogata Y 2003
{\it Variational principle for non-equilibrium steady states of the XX
model}
Rev. Math. Phys. 15 905

\bibitem{MK} Mikeska H-J and Kolezhuk A K 2004
{\it One-dimensional magnetism} in
Schollw\"ock U, Richter J, Farnell D J J, and Bishop R F (Ed.)
{\it Quantum Magnetism} Lect. Notes Phys. 645 1 (Springer)

\bibitem{O} Ogata Y 2002
{\it Nonequilibrium properties in the transverse XX chain}
Phys. Rev. E 66 016135 


\bibitem{R} Ruelle D 2001
{\it Entropy production in quantum spin systems}
Commun. Math. Phys. 224 3

\end{thebibliography}
\end{document}